\documentclass[12pt,a4paper]{amsart}
\usepackage{amsmath,amsthm,latexsym,amssymb,amsfonts,amsbsy,mathrsfs,braket}
\usepackage[mathscr]{eucal}
\usepackage{graphicx,multirow,tabularx}
\usepackage{amsfonts,delarray}
\usepackage{algorithm,algpseudocode}
\usepackage{bm}
\usepackage{epstopdf}

\usepackage{color}

\usepackage[comma]{natbib}

\numberwithin{equation}{section}  

\usepackage{subfigure,times,rotating}
\usepackage{geometry}

 \rm
 \rm
 \rm
\DeclareMathOperator*{\argmin}{arg\,min}

\title[Deconvolving PET with FPCA]{A functional approach to deconvolve dynamic neuroimaging data}

\author[Jiang, Aston \& Wang]{C. R. Jiang$^{1}$, J. A. D. Aston$^{2}$ and J. L. Wang${^3}$\\$^{1}$Institute of Statistical Science, Academia Sinica\\$^{2}$Statistical Laboratory, University of Cambridge\\$^{3}$Dept. of Statistics, University of California, Davis}

\address{
\begin{tabular}{ll}
\\
Address for Correspondence:\\
John Aston\\Statistical Laboratory&Tel: \texttt{+44-1223-766535}\\
DPMMS&Email: \texttt{j.aston@statslab.cam.ac.uk}\\
University of Cambridge\\
UK\\
\end{tabular}
}


\date{\today}

\begin{document}

\begin{abstract}
Positron Emission Tomography (PET) is an imaging technique which can be used to investigate chemical changes in human biological processes such as cancer development or neurochemical reactions. Most dynamic PET scans are currently analyzed based on the assumption that linear first order kinetics can be used to adequately describe the system under observation. However, there has recently been strong evidence that this is not the case. In order to provide an analysis of PET data which is free from this compartmental assumption, we propose a nonparametric deconvolution and analysis model for dynamic PET data based on functional principal component analysis. This yields flexibility in the possible deconvolved functions while still performing well when a linear compartmental model setup is the true data generating mechanism. As the deconvolution needs to be performed on only a relative small number of basis functions rather than voxel by voxel in the entire 3-D volume, the methodology is both robust to typical brain imaging noise levels while also being computationally efficient. The new methodology is investigated through simulations in both 1-D functions and 2-D images and also applied to a neuroimaging study whose goal is the quantification of opioid receptor concentration in the brain.

\end{abstract}

\maketitle

\newpage
\section{Introduction}\label{s:intro}

Positron Emission Tomography (PET) is an \textit{in-vivo} neuroimaging technique for studying biological processes in humans. It is almost unique amongst the major neuroimaging modalities, in that it can be used to study neurochemical concentrations and associated changes in a quantifiable way. PET works on the principle of using an injected radioactive tracer compound specifically designed for the biological process of interest and tracking its presence throughout the target organ through the emitted radiation of the radioactively decaying compound. It is a quantitative technique, as opposed to say functional Magnetic Resonance Imaging (fMRI), in that the amount of radiochemical injected can be used to establish the concentrations present in the target organs. This has led it to be almost universally used in the diagnosis of certain cancers through fluorodeoxyglucose (FDG) PET scans, which as a surrogate for glucose, can be used to target tissues with high metabolic rates, something characteristic of cancer cells. Indeed, it is not only used for cancer diagnosis and localization in the brain, but also throughout the body \citep{Gamb:02,Hsieh:12}.

In addition to diagnostic and clinical usage, PET also can be used to investigate neurochemical processes to help further understanding of the brain. Individual neurochemical transmitter systems can be targeted through the design of radiotracers that mimic the behavior of these chemicals, but which being radioactive can be traced by the PET camera. As might be imagined, this involves considerable complex radiochemistry to design suitable radiotracers. However, there are now many tracers available to target systems in addition to metabolism such as the dopamine system \citep{Wagner:83}, the serotonergic system \citep{Drevets:99} and the opioid receptor system \citep{Jones:04}. Indeed, it is the last of these, the opioid system, that is the motivation for this work. The opioid system controls the brain's reaction to pain \citep{pasternak:1993}, and has been associated with a number of conditions and diseases including changes in emotional responses \citep{filliol:2000}, addiction \citep{wise:1996} and Alzheimer's disease \citep{jansen:1990}. The data which will be analyzed later in this paper is taken from part of a large study on the role of the opioid system in Epilespy. It is of great interest to get accurate quantifiable estimates of opioid receptor concentrations and densities throughout the brain in normal subjects as a precursor to understanding the role of receptor changes in disease diagnosis, prognosis and treatment.

PET scans, in a similar way to fMRI scans, consist of 3-D volumes of data recorded over time, leading to large data sets with time courses from millions of spatial locations (voxels). These 3-D volumes are tomographic reconstructions. However, the reconstruction process will not be the focus here, as typical PET data is reconstructed using reprojection algorithms \citep{KinaR:89} and then analyzed as if this reconstructed data was actually measured directly. In order to facilitate usage by practitioners and comparison with the most commonly used approaches in scientific and clinical application, we will also focus our analysis on these reconstructed data. However, in principal, with suitable modification, the methods introduced in this paper could be incorporated into reconstruction in a similar way to compartmental analysis (eg \cite{WangQi:09}).

The time courses associated with PET data are characteristically non-linear in that, being associated with chemical reactions, they are routinely modelled as coming from ordinary differential equation (ODE) systems, where first order linear kinetics can be used to model the data \citep{Gunn01}. These kinetics are routinely associated with compartmental models, which consist of abstract compartments within each voxel. The transfer of material from one compartment to another is assumed to follow a first order ODE. For more information on compartmental models see \citet{godfrey1983compartmental}. However, there is increasing evidence both from biological experiments and statistical analysis that such models are not adequate for the data \citep{OSullivan:09}, not least because each voxel represents an inhomogeneous mixture of cells leading to a mixture of compartmental processes (assuming the compartmental assumption is even made). In addition, fitting methods which are stable for large numbers of voxels, such as non-negative least squares, tend to have parameter dependent bias, while methods such as non-linear least squares tend to be somewhat unstable \citep{PengAGLA:08}. In order to account for the model discrepancies while still maintaining a robust approach to model fitting, this paper explores a nonparametric deconvolution model for PET analysis. The input (through the blood flow to the brain) can be measured online to all extents and purposes continuously with virtually no measurement error relative to the error in the measured voxelwise PET data, as the sampling on the measured radioactivity of the blood is done outside the body using a sensitive blood monitor via an arterial canula to produce a smooth continuous input curve \citep{Lamm:91}. This allows one of the functions in the deconvolution to be known (i.e. this is not a blind deconvolution problem), but the inherent difficulties of deconvolving the noisy measured output function are all still present. We present a new methodology for deconvolution and analysis of data. This new methodology works when there are multiple observations of the convolved functions, and can also be used when the functions are possibly dependent on a covariate.

The analysis is based on using functional principal component analysis (FPCA). Our methodology involves a presmoothing step to reconstruct the image, followed by a deconvolution step to recover the impulse response function. Presmoothing decreases the noise in the data, hence potential biases in further analysis. In parametric non-linear models such as compartmental models, bias can be noise dependent \citep{PengAGLA:08}, while here the errors in the observed functions are somewhat similar to those in measurement error models, which yield biases in traditional regression analysis. The presmoothing also produces functions that are smoother than the original data, making subsequent deconvolution easier.  As for the deconvolution approach, ours  differs from traditional deconvolution methods and has inherent computational advantages in that we treat the sample of dynamic PET data on all voxels as functional data, and apply FPCA to reduce the dimension of the data, so that the deconvolution  only needs to be performed on the mean and eigenfunctions of the data.   This has substantial computational advantages as while there are millions of spatial voxel locations, often only a few basis functions in the FPCA basis are needed to adequately describe the temporal curves, requiring only a very small number of actual deconvolutions to be performed.
Moreover, it is not the actual deconvolutions that are the focus of the PET study. Of primary interest in many PET studies is the volume of distribution, $V_T$, the integral of the impulse response function of the system at each voxel. Under various biological assumptions, $V_T$ can be used to determine the receptor density of the underlying neurotransmitter \citep{RLNomen}. As advocated in \cite{OSullivan:09}, this will be approximated by the integral of the deconvolved response function generated from the observed data, which in itself is a more meaningful measure as it is less dependent on the particular compartmental model fit assumed.

The paper proceeds as follows. In the next section, the moderately general methodology, inspired by PET data, is introduced for deconvolution of multiply observed functions through the use of FPCA. In section \ref{s:sims}, the methods are assessed through simulation, not only on 1-D functions, but also on moderately realistic 2-D image slices where both spatial correlations and non-homogeneous noise models, typical of those found in PET studies, are used. In section \ref{s:data}, the methods are applied to measured [$^{11}$C]-diprenorphine scans taken from healthy volunteers and are used to provide voxelwise quantification of receptor concentration without resorting to compartmental assumptions. The final section discusses some of the possible extensions of this work.

\section{Methodology}\label{s:methodology}

Let $C_i(t)$ be the concentration curve of voxel $i$ in PET analysis.  The conventional assumption is that
\begin{equation}\label{eq:concen}
C_i(t)=I(t)\otimes M_i(t)= \int_0^t I(t-s)M_i(s)ds,
\end{equation}
where  $I(t)$ is a known input function and $M_i(t)$ is the unknown impulse response function (IRF) of voxel $i$.  In reality, $C_i(t)$ is not observed, but rather, a noise contaminated version of $C_i(t)\exp(-\lambda t)$  is observed \citep{AstoGWME:00} at discrete time points, $t=t_1,\ldots,t_p$ where $\lambda$ is the known decay constant of the radioisotope (in the case of $^{11}C$, this is $5.663 \times 10^{-4} s^{-1}$.).   Hence the observations for the $i$th voxel are $Y_{ij}= C_i(t_j)\exp(-\lambda t_j) + \varepsilon_{ij}$ , where $\varepsilon_{ij}$ are independent noise for $i=1,\ldots, n$. Here, the independence assumption on the errors can be largely justified on the basis of the independent Poisson decay nature of radioactivity \citep{Carson:85}.

The goal of PET analysis is to estimate the volume of distribution ($V_T$) at each voxel $i$, which is $V_T(i)=\int_0^\tau M_i(t)dt$, where $\tau$ is the end of the experimental time.   In order to estimate $V_T$, it is necessary to estimate the IRF $M_i(t)$ through deconvolution. As we are using a nonparametric estimator in the deconvolution, it is not possible to extrapolate the $V_T$ to infinity (as this would require a parametric model), but this finite truncated version could well be preferred in many situations \citep{OSullivan:09}, particularly given the known difficulties of function extrapolation.

\subsection{Spatial Curve Pre-regularization}

With the presence of noise in the  output data  $Y_{ij}$, our first step is to reconstruct
$C_i(t)\exp(-\lambda t)$ for all voxels.  Instead of handling these temporal curves voxel by voxel, we borrow spatial information from all voxels by applying a spatially adapted smoother to $Y_{ij}$ across all time points ($t$) and spatial/voxel locations, denoted as $Z_i$ for the $i$th voxel.   Depending on the dimension of the image,  a three (for 2D images) or four (for 3D images) dimensional smoother is used  to reconstruct the latent signals. For the PET data in Section \ref{s:data}, $Z_i$ is three-dimensional, so a four-dimensional smoother is employed. This may seem a formidable task, given the large amount of available data (32 time points and 150,784 brain voxels), but it is feasible if one adopts an computationally efficient approach.

Specifically, the smoothed estimate of $C_i(t)\exp(-\lambda t)$ at voxel $i$ is:
\begin{align}\label{eq:pp}
\hat{Y}_i(t) & = \hat{b}_{i,0}(t), \text{ where}\\ \nonumber
\hat{\bm b}_i(t) & = \arg\min_{\bm b} \sum_{k=1 }^n \sum_{j=1}^p K_{kj,h}(z_i, t )\left\{Y_{kj} - b_{i,0} - \sum_{\ell=1}^3 b_{i,\ell}(z_{i\ell}-z_{k\ell})-b_{i,4}(t-t_j) \right\}^2,
\end{align}
and $K_{kj,h}(z_i,t)= \frac{1}{\beta \hat{h}_T(t)h_{z_1}h_{z_2}h_{z_3}} K(\frac{z_{i1}-z_{k1}}{h_{z_1}},\frac{z_{i2}-z_{k2}}{h_{z_2}},\frac{z_{i3}-z_{k3}}{h_{z_3}},\frac{t-t_j}{\beta \hat{h}_T(t)})$ is a four-dimensional kernel function (an Epanechnikov kernel was used in the data analysis). Note that constant bandwidths are employed for spatial coordinates (in the application, one bandwidth is chosen for all three dimensions), but an adaptive local bandwidth for the time dimension is applied (see section \ref{s:vb} for details).  The reconstructed concentration function for $C_i(t)$ is
\begin{equation}  \label{eq:Chat}
\hat{C}_i (t) = \hat{Y}_i(t)\exp(\lambda t).
\end{equation}



If a kernel estimator is chosen, a product kernel can be applied to save computational time, which is equivalent to smoothing each coordinate of time and space sequentially.

\subsection{Variable Bandwidth}\label{s:vb}
In most PET analysis, particularly in the spatial domain, smoothing is based on heuristic assessments determined by the individual researcher. Here we propose to use data driven methods to select the bandwidth choices. A constant bandwidth is suitable for the spatial coordinates as the covariance structure, while subtly changing across the image, does not vary substantially. However, in the time coordinate, due to the denser measurements at the beginning of the time period and the sharp peak near the left boundary, a non-constant bandwidth is required. To retain the peak without compromising the performance at other temporal locations, a locally adaptive bandwidth function is recommended and applied in our analysis. Essentially, a smaller bandwidth is preferred near the peak location, while larger bandwidths are used near the right boundary where the curve is relatively flat. This is also consistent with the fact that the noise in PET data can be crudely seen as being Poisson distributed due to the radio labeled nature of the data \citep{Carson:85}.

\begin{figure}[h]
\caption{The resulting locally adaptive bandwidth for PET time-course data.}\label{f:bdf}
\begin{center}
\includegraphics[scale=0.4,angle=270]{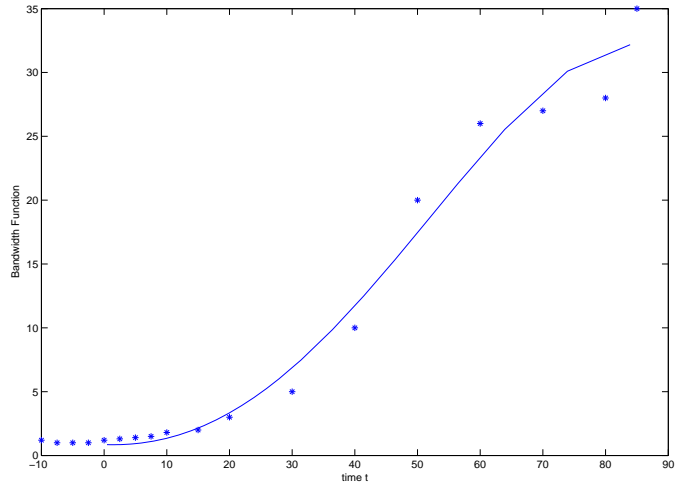}
\end{center}
\end{figure}

We undertook the following pragmatic approach to design such a bandwidth function. First, a number of time locations ($n_b$) $(t_{(1)},\ldots,t_{(n_b)})$, where the time-course data were observed, were selected (we used $n_b = 13$ in the application, which was approximately 1/3 of the time points in the time course). At each location, the bandwidth $h_T(t)$ at location $t$ was chosen such that the interval $[t-h_T(t), t+h_T(t)]$ contains at least four observations. Further, boundary correction was employed to ensure the resulting bandwidth function was positive when $t$ was close to zero. A fourth order polynomial was applied to the pair set $\{(h_T(t_{(i)}),t_{(i)})| i=1,\ldots,n_b\}$ to obtain a smooth bandwidth function. The resulting bandwidth function $\hat{h}_T(t)$ (shown in Figure \ref{f:bdf}) was further multiplied by a constant $\beta$. The constant $\beta$ serves to facilitate calibration of the final local bandwidths, because the choice of local bandwidths for $h_T(t)$ was subjective, and thus $\beta$, which was determined by cross-validation, allowed this subjective choice to be adapted to the data. This form of bandwidth selection has been shown to work well in previous studies on smoothing prior to parametric compartmental modelling \citep{JianAW:09}. 

While several tuning parameters need to be chosen for the analysis, this is not uncommon in PET, as data is usually smoothed to increase the signal to noise ratio or to facilitate population studies. However, as mentioned above, most analyses use the default settings of whatever software package is being used, while we here prefer to determine a good choice of bandwidth through cross validation.

\subsection{Deconvolution Based on FPCA}

With the concentration curve reconstructed for each voxel, one can perform deconvolution voxelwise to recover the IRF. However, attempting to perform automated deconvolution over such a large number  of functions is inherently problematic and computationally costly. Alternatively, we take the viewpoint that the concentration curves are random curves, a.k.a. functional data (\cite{RamsS:05}), so a functional approach can be employed to model these curves.  Since
convolution is a linear operator, it is advantageous to adopt a linear mixed-effects model approach to represent these functional data.  Since we do not assume that the shapes of the IRFs are known a-priori, a nonparametric basis function is the preferred choice and  we adopt parsimonious basis functions through principal component analysis.

Principal component analysis is a popular dimension reduction approach for multivariate data and has been extended to functional data that are in the form of random curves and termed FPCA.  Many different FPCA approaches have been developed, such as by \cite*{DauxPR:82}, \cite{RiceS:91},  \cite*{BoenF:00},  \cite{Card:00}, and \cite{YaoMW:05:1}.  We adopt a similar approach as  \cite{YaoMW:05:1}  but with a slightly different model that was advocated in \cite{JianAW:09} for PET time course data.  Specifically, a multiplicative random effects model was proposed there, motivated by the likely randomness in chemical rates (induced by spatially varying neurochemical receptor densities, for example) leading to multiplicative changes in the curves.  Thus, we adopt the following modified Karhunen-Lo\`{e}ve decomposition of $C_i(t)$, which includes  an additional random effect term $A_{i0}$  on the mean function:
\begin{equation}\label{eq:mfpca3}
C_i(t)=A_{i0}\mu(t)+\sum_{k}A_{ik}\phi_k(t),
\end{equation}
where $\mu(t)=E\{C_i(t)\}$ is the mean function, $E(A_{i0})=1$, $\phi_k(t)$ are the eigenfunctions of  the covariance of $C_i(t) - A_{i0} \mu(t)$ with the corresponding non-increasing eigenvalue $\zeta_k$, and $A_{ik}$ is the $k$-th functional principal component score.

In general deconvolution is an ill-posed problem. However, due to the positivity of the input function $I(t)$, (\ref{eq:concen}) and (\ref{eq:mfpca3}) imply that
\begin{equation}\label{eq:Mt}
M_i(t) = A_{i0} \mu^d(t) + \sum_{k}A_{ik}\phi^d_k(t),
\end{equation}
where $\mu(t) = I(t)\otimes \mu^d(t)$ and $\phi_k(t) = I(t)\otimes \phi^d_k(t)$. Therefore, deconvolution only has to be performed on the mean function and the likely small number of eigenfunctions needed to give a good representation of the data.  This has considerable computational savings compared to performing it on hundreds of thousands of spatial voxels,  and is one of the main advantages of our approach. It should, however, be noted at this point that $\mu^d(t)$, and $\phi^d(t)$ do not necessarily form an eigendecomposition of $M_i(t)$ but are rather a basis of the deconvolved space.


To perform the deconvolution, we consider the following strategy, which will be illustrated on  $\mu(t)$.  Suppose that $\mu(t)$ or an estimate of it is available at times $s_0, s_1, \ldots, s_m$, where $s_0=0$ and $s_m=\tau$.   Let $\bm{\mu}^T= (\mu(s_1), \ldots, \mu(s_m))$.    When $m$ is large,
$\bm{\mu} \approx \mathbb{A}\bm{\mu}^d$,
where
\begin{equation}\label{eq:convapp}
\mathbb{A} =
\begin{pmatrix}
  I(s_1)\frac{s_1}{2} & 0          & 0 & \ldots & 0 \\
  I(s_2)\frac{s_1}{2} & I(s_2-s_1)\frac{s_2}{2} & 0 & \ldots & 0 \\
  \vdots & \vdots & \ddots & \vdots \\
  I(s_m)\frac{s_1}{2} & I(s_m-s_1)\frac{s_2}{2} & I(s_m-s_2)\frac{s_3-s_1}{2} & \vdots & I(s_m-s_{m-1})\frac{s_m-s_{m-2}}{2} \\
\end{pmatrix},
\end{equation}
and $\bm{\mu}^{d} = (\mu^d(s_0), \ldots, \mu^d(s_{m-1}))^T$. The matrix $\mathbb{A}$ can be seen as a linear discretisation of the convolution integral. Therefore, we can obtain an estimate of $\bm{\mu}^d$ by
\begin{equation}\label{eq:decov}
\widehat{\bm{\mu}^d}= \argmin_{\bm{\mu}^d} \|\bm{\mu}-\mathbb{A}\bm{\mu}^d\|^2.
\end{equation}
This allows the deconvolution procedure to be framed as a linear regression problem, allowing the use of the usual standard least squares formulation. In the measured data analysis and simulations in the next sections, we interpolated the smoothed PET time courses to $m=250$ to balance computational complexity with discretisation error.

This is of course not the only possible deconvolution strategy that could be used, and many others exist in the literature, including spline based deconvolution as used in \cite{OSullivan:09,OSullivan:14}. However, it is very simple and computationally efficient to implement, and as will be seen in the simulations produces reasonable estimates of the deconvolved curves.

\subsection{Estimation of FPCA}   Since the mean function $\mu(t)$ and eigenfunctions $\phi_k(t)$ associated with the concentration function  $C_i(t)$ are unknown, they need to be estimated first before one can  implement the deconvolution in  (\ref{eq:decov}).

\noindent  {\bf Estimation of  the mean function} $\mathbf{\mu(t).}$  \  One could use the mean function of the reconstructed $\hat{C}_i(t)$ in (\ref{eq:Chat}). However, as $\hat{C}_i(t)$ results from smoothing, the bias inherited at this step in the reconstruction leads to a biased estimate of $\mu$. We thus estimate  $\mu$ through the sample mean of $Y_{ij}$.  Let   $Y_{\cdot  j} = \frac{1}{n} \sum_{i=1}^n Y_{ij} $  be  the cross-sectional mean (without any smoothing) of the observed data $Y_{ij}$  at time $t_j$.  The estimate for $\mu(t)$ is
\begin{eqnarray}  \label{eq:Muhat}
\hat{\mu}(t) &=& Y_{\cdot  j} \exp(\lambda t),  \   \text{ for} \ t=t_j, \\  \nonumber
&=& \mbox{the linear \ interpolated \  value \ of } \hat{\mu}(t_k) \  \mbox{and} \ \hat{\mu}(t_{k+1}),  \ \mbox{ for}  \  t_k < t < t_{k+1}.
\end{eqnarray}
The resulting linearly interpolated estimate $\hat{\mu} (t)$ is unbiased at $t=t_j$ for all $j=1, \ldots, p,$ and has a smaller bias at other $t$ than the mean of $\hat{C}_i(t)$. Of course, other interpolating schemes such as cubic spline interpolation could be substituted at this point, but linear interpolation is faster and seems to work well for PET data.

\noindent  {\bf Estimation of $A_{i0}$.}  \    The estimate of the multiplicative coefficient at voxel $i$ is
\begin{equation}  \label{eq:Ai}
\hat{A}_{i0} = \int_0^{\tau} \hat{C}_i(t)\hat{\mu}(t)dt / \int_0^{\tau}\left(\hat{\mu}(t)\right)^2dt,
\end{equation}
where $\hat{\mu}(t)$ is the estimate of the mean function $\mu$ from (\ref{eq:Muhat}) and $\hat{C}_i(t)$  is from (\ref{eq:Chat}). It should, of course, be noted here that the resulting $A_{i0}$ will not necessarily have the usual property of having expectation one. This results from estimating the $A_{i0}$ from the smoothed data while the mean function is derived from the unsmoothed data. However, in practice the difference between the two is small, and considerably less variable estimation results from using smoothed data to estimate $A_{i0}$ (a classic bias-variance trade-off).

\noindent  {\bf Estimation of the eigenfunctions} $\mathbf{\phi_k}$   and {\bf principal component scores.} \ We estimate the covariance function by the sample covariance of $\hat{C}_i(t_j)- \hat{A}_{i0}\hat{\mu}(t_j)$, where $\hat{C}_i$ is from (\ref{eq:Chat}),  $\hat{A}_{i0}$ is from (\ref{eq:Ai}), and $\hat{\mu}$ is from  (\ref{eq:Muhat}).   Specifically,
\begin{equation}  \label{eq:Gammahat}
\hat{\Gamma}(t_j,t_k) = \frac{1}{n}\sum_{i=1}^n \{\hat{C}_i(t_j)- \hat{A}_{i0}\hat{\mu}(t_j)\}\{ \hat{C}_i(t_k)- \hat{A}_{i0}\hat{\mu}(t_k)\}
\end{equation}
for $1\leq j, k \leq p$. Once the covariance is obtained, the eigenfunctions can be estimated by solving the eigen-equations at a dense grid.  Let $\hat{\phi}_k(t)$ be the estimate of $\phi_k(t)$,  The principal component scores $A_{ik}$ can be estimated by
\begin{equation} \label{eq:Akhat}
\hat{A}_{ik} = \int_0^{\tau}  \{\hat{C}_i(t)- \hat{A}_{i0} \hat{\mu}(t)\} \  \hat{\phi}_k(t) \ dt.
\end{equation}

\noindent  {\bf Number of components.} The number of eigenfunctions $L$ for voxel $i$ is selected by
      \begin{equation}\label{eq:R2}
      R^2(i,L)  = 1 - \frac{\text{var}\left\{Y_i(t)-\hat{C}_i(t,L)\exp(-\lambda t)\right\}}{\text{var}\{Y_i(t)\}},
      \end{equation}
      where $\hat{C}_i(t,L)  = \hat{A}_{i0}\hat{\mu}(t) + \sum_{k=1}^L \hat{A}_{ik} \hat{\phi}_k(t).$
The above $R^2$ is an ad hoc measure for the goodness of fit, but provides a useful summary of how much additional information is gained by adding a further eigenfunction. For the simulation and data analysis in later sections, we adopted a simple rule to select $L$ by setting $L=k$ when  $R^2(i,k+1)-R^2(i,k)<0.025$.

After the number of components $L$ is selected by $R^2$, the IRFs can be reconstructed through (\ref{eq:Mt}) and its associated $V_T$ can be estimated by integration of the IRF.  Specifically, at the $i$th voxel, these estimates are:
    \begin{align} \label{eq:Mhat}
      \hat{M}_i(t,L) & = \hat{A}_{i0}\hat{\mu}^d(t) + \sum_{k=1}^L \hat{A}_{ik} \hat{\phi}^d_k(t),\\ \label{eq:VDhat}
       \widehat{V_T}(i,L) & = \hat{A}_{i0} \int_0^\tau\hat{\mu}^d(t)dt + \sum_{k=1}^L \hat{A}_{ik} \int_0^\tau\hat{\phi}^d_k(t)dt.
      \end{align}
The details of our approach are summarized in Algorithm \ref{alg:deconv}.

\begin{algorithm*}[htb]
  \caption{Deconvolving PET with FPCA.}
  \label{alg:deconv}
  \begin{algorithmic}[1]
    \Require
      The set of PET time course data $\{Y_{ij};i=1,\ldots,n, j=1,\ldots,p\}$;
    \Ensure
      The $V_T$ of each voxel $i$;
    \State \textbf{Pre-Processing}
    \begin{enumerate}
      \item[] Pre-process $Y_{ij}$ by (\ref{eq:pp}) and
           reconstruct $C_i(t)$ by (\ref{eq:Chat});
    \end{enumerate}
    \label{code:pp}

    \State \textbf{FPCA}
    \begin{enumerate}
    \item Estimate the mean function $\mu(t)$ by (\ref{eq:Muhat});
    \item Estimate the covariance function $\Gamma(s,t)$ by (\ref{eq:Gammahat});
    \item Perform eigen-decomposition on $\hat{\Gamma}(s,t)$ to obtain $\hat{\phi}_k(t)$;
    \item[] Calculate the PC scores by (\ref{eq:Akhat});
    \item Select the number of eigenfunctions $L$ by (\ref{eq:R2});
    \end{enumerate}

    \State \textbf{Deconvolution}
    \begin{enumerate}
    \item[] Perform deconvolutions on $\hat{\mu}(t)$ and $\{\hat{\phi}_1(t),\ldots,\hat{\phi}_L(t)\}$ via (\ref{eq:decov});
    \end{enumerate}

    \label{code:lls}
    \State Calculate $V_T$ for each voxel by (\ref{eq:VDhat});
  \end{algorithmic}
\end{algorithm*}

\section{Simulation Studies}\label{s:sims}

The proposed methodology is firstly evaluated on simulated data, both on 1-D functions and then in more realistic image settings using 2-D image phantoms.

\subsection{One Dimensional Function Simulations}

\begin{figure}[h]

\caption{First eight observed curves of the 200 curves in the 1D simulation.}\label{sim0_dat}
\begin{center}
\includegraphics[width=\textwidth]{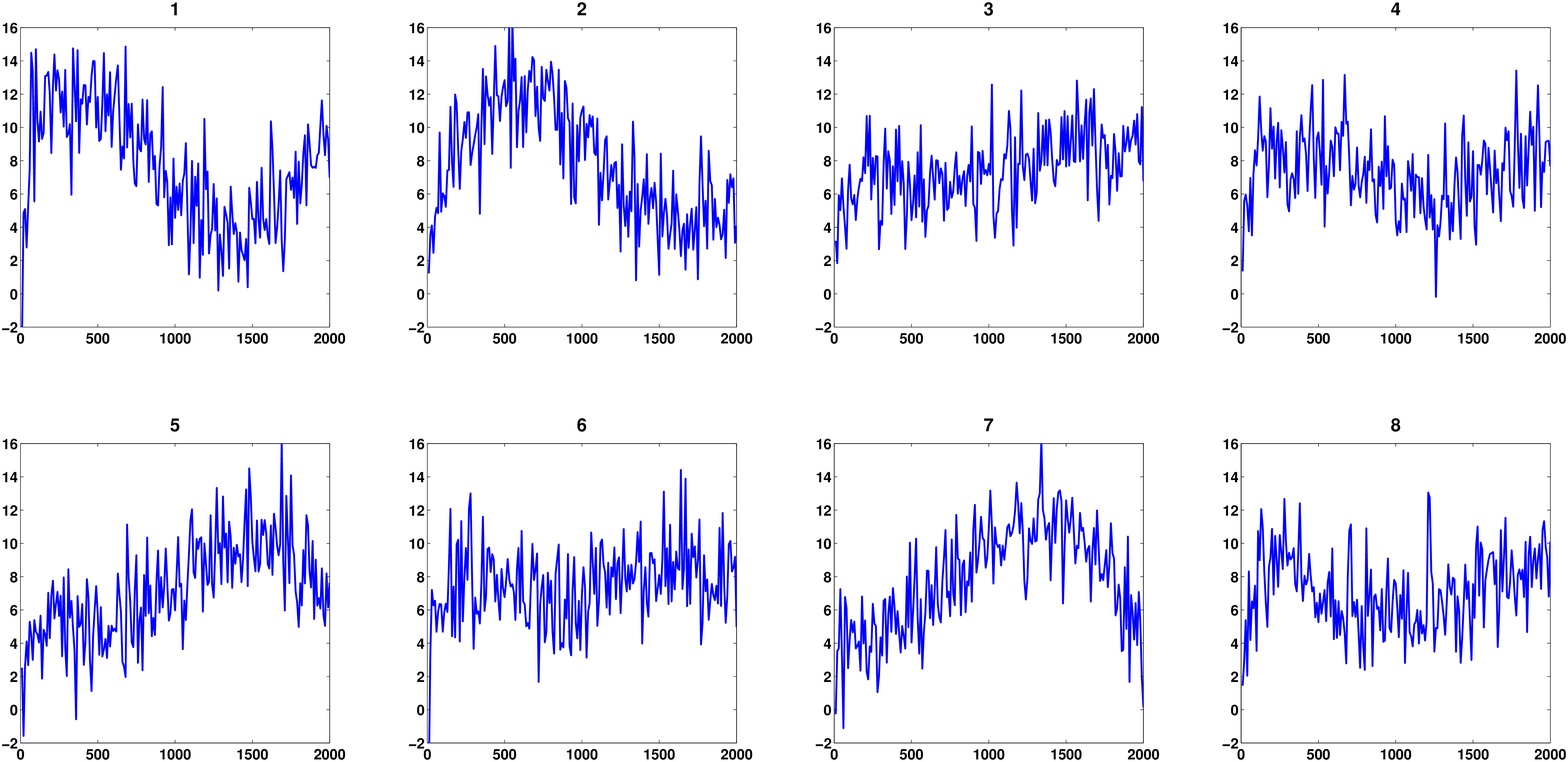}
\end{center}
\caption{Estimated deconvolved functions and true target function in 1D simulation corresponding to the curves in Figure \ref{sim0_dat} along with the pointwise MSE for each method.}\label{sim0_fig}
\begin{center}
\includegraphics[scale=0.2]{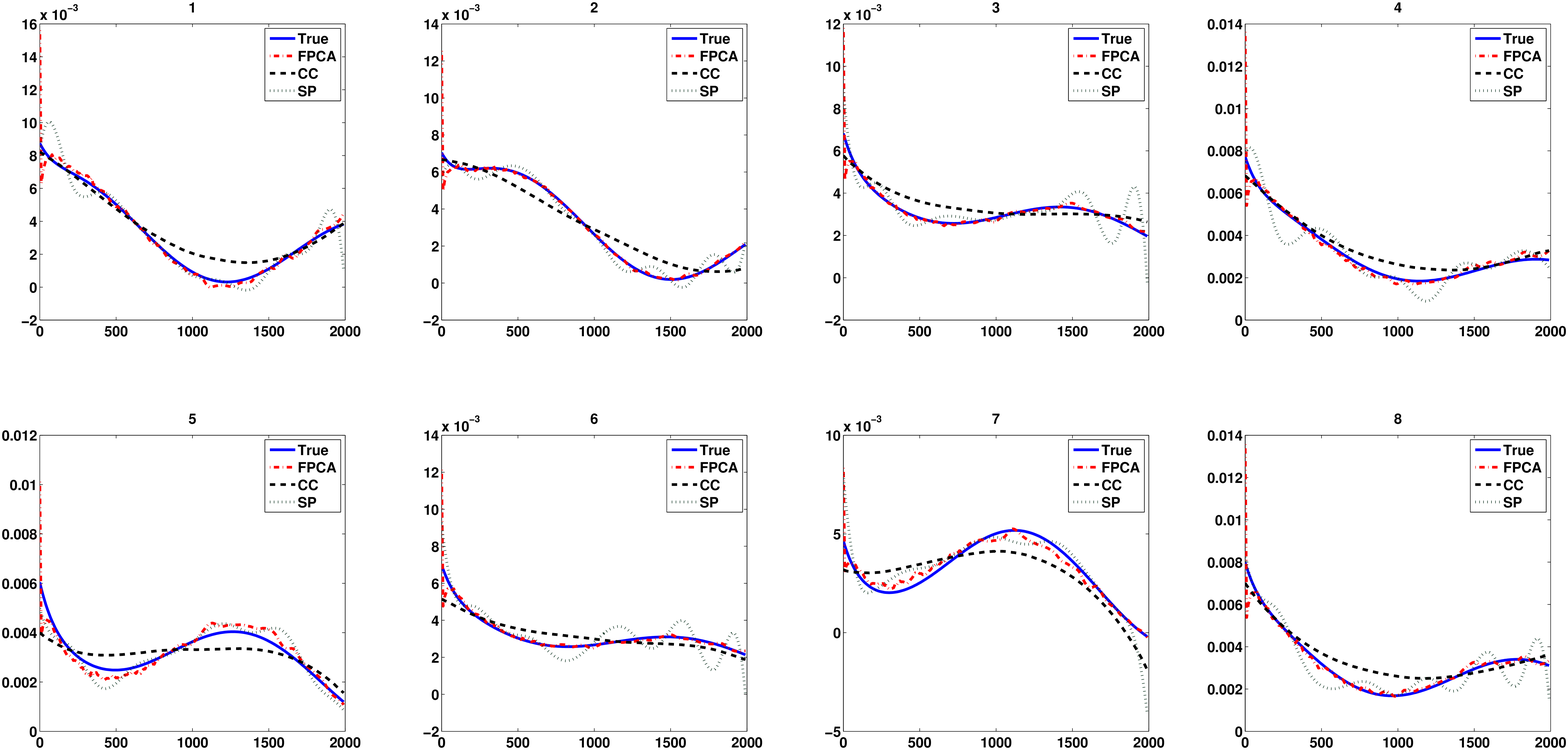}
\includegraphics[scale=0.4]{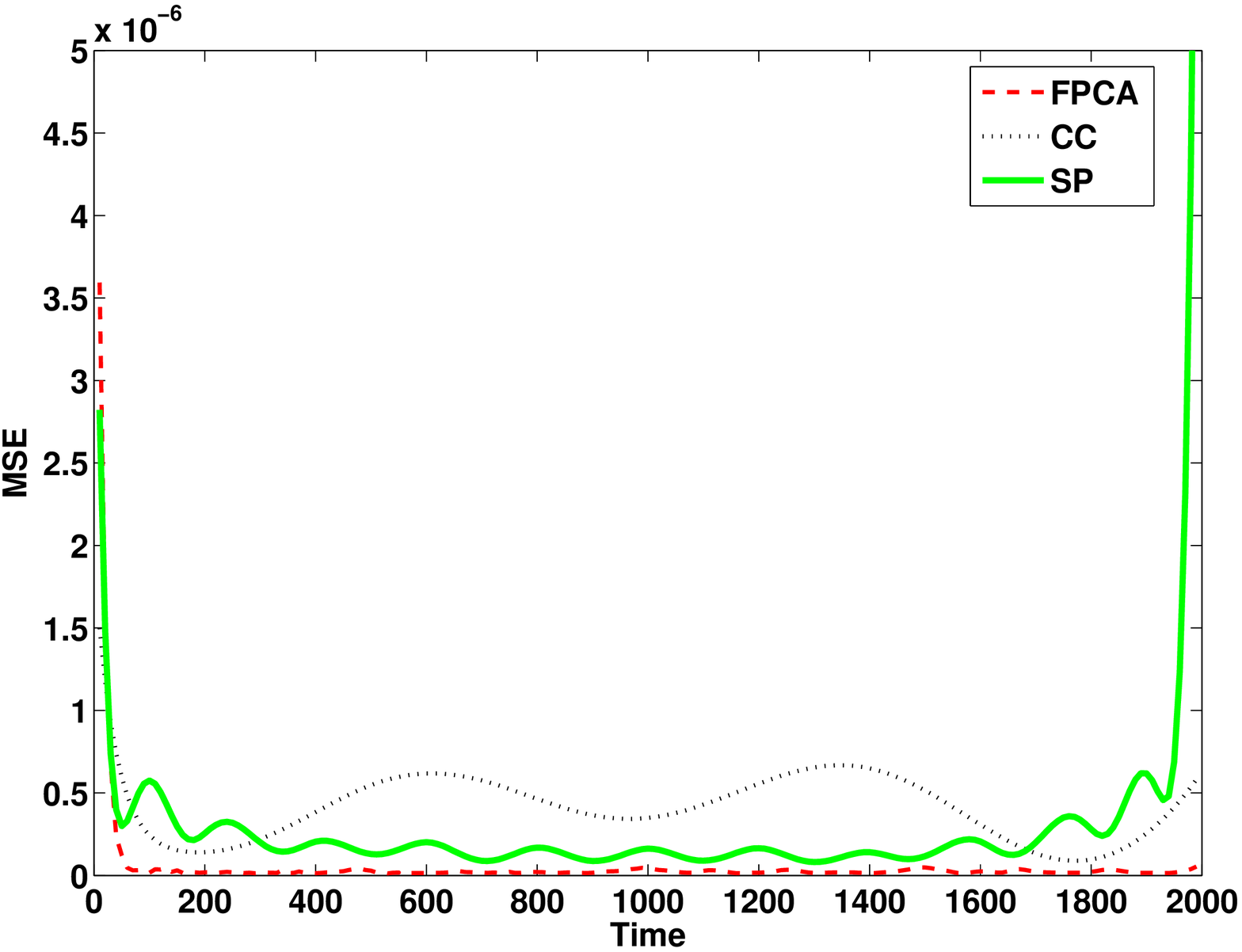}
\end{center}

\end{figure}

Before showing our formal image simulation studies, we would like to demonstrate the use of the deconvolution strategy via FPCA and assess how well it works for general functional data. First, we assume the target functions (IRFs in PET imaging data) can be represented as
$$
M_i(t) = \mu_M(t) + \sum_{k=1}^2 B_{ik} \psi_k(t),
$$
where $B_{ik}$ are random variables and $\psi_k(t)$ are basis functions. Specifically, we simulate the mean function $\mu_M(t) = .0049\exp(-.0005t)+.0018\exp(-.0112t)$ and  $\psi_1(t) = c_1\sin(2\pi t/2000)$ and $\psi_2(t) = c_2\cos(2\pi t/2000),$ where $c_1$ and $c_2$ are constants which normalise the basis functions in the $L_2-$norm. Specifically,
\begin{equation*}
c_1 = \frac{1}{\sqrt{\int_0^{2000} \sin^2(2\pi t/2000)dt }} \text{ \hspace{1mm}  and \hspace{1mm} } c_2 = \frac{1}{\sqrt{\int_0^{2000} \cos^2(2\pi t/2000)dt}}.
\end{equation*}
Also, the random coefficients ($B_{i1}$ and $B_{i2}$) for the basis functions are generated from $N(0,.1^2)$ and $N(0,.05)$ respectively. The random functions are then convolved with an arterial input function taken from the  [$^{11}$C]-diprenorphine study of the next section truncated at $2000$ seconds.
The data observed are further contaminated with independent measurement errors at the observation times,
$$
Y_i(t) = I(t)\otimes M_i(t) + \varepsilon,
$$
where $\varepsilon\sim N(0,2^2)$ and where notationally we assume that the errors are only present at the observations, not over the entire continuum. This toy example contains 200 curves with observations made at 200 equally spaced time points and the first eight observed noisy curves are shown in Figure \ref{sim0_dat}. The MATLAB package PACE \citep{YaoMW:05:1} was applied to obtain the mean function and eigenfunctions for the observed functions. Figure \ref{sim0_fig} indicates that our deconvolution strategy via FPCA performs very well for regular functional data. As the FPCA utilizes information across all curves, this improves the deconvolution.

In addition to the FPCA approach, we compared several other approaches. \cite{OSullivan:09} proposed an approach which worked well for region of interest (ROI) data in FDG. It was based on a spline deconvolution of the response function. We implemented an analogous spline based deconvolution (SP), using a weight function suitably chosen for the simulations. We also examined deconvolution as used within the FPCA procedure (CC) on a curve-by-curve basis. However, both the spline deconvolution and the CC deconvolution of the same data is much less accurate (see Figure \ref{sim0_fig}) with a 10 fold increase in mean integrated squared error (MISE) between the FPCA approach and the CC approach (MISE, FPCA: $0.0879 \times 10^{-3}$, CC: $0.8160 \times 10^{-3}$, SP: $0.5045 \times 10^{-3}$), with the spline approach performing slightly better than the CC approach except at the boundaries, but still less well than the FPCA approach. Similar results (not shown) were obtained if the input function was replaced by a known function, such as a scaled gamma function, rather than the input function from the measured data.

\subsection{Image Based Simulations}

In the context of using PET data, the structure of the data is considerably more complex than was used in the 1-D function simulations above. In particular, there is considerable spatial correlation in the measured data due to both the inherent underlying biological physiology as well as the blurring induced by the resolution of the PET camera. Theoretically, weak dependence of this sort is not an issue for FPCA \citep{HormK:10,HormK:13}. However, from a practical point of view, the performance of the proposed methodology is now assessed in light of these factors.

\subsubsection{Simulated Data Generation}

\begin{figure}[h]
\caption{Phantom Image: each of the five different regions used in the simulations are indicated.}\label{phantom_img}
\begin{center}
\includegraphics[scale=0.4]{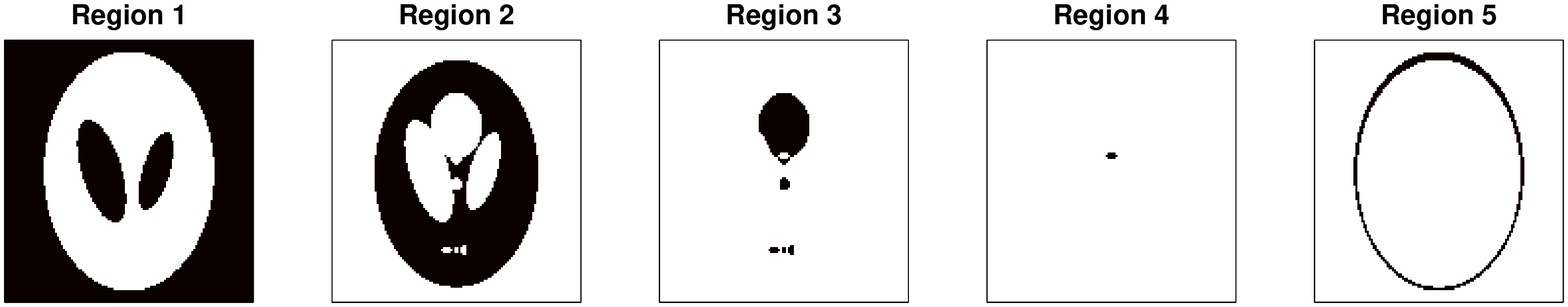}
\end{center}
\end{figure}

In order to assess the effect of different regions, simulations was performed using a brain phantom (Shepp-Vardi phantom, $128 \times 128$ pixels) with five different regions of varying sizes (Figure \ref{phantom_img}). Different signals were placed in each of the regions based on random parameters which also depended on the type of simulation being performed. The level of $V_T$ randomness in each region was about $6.5\%$ roughly equivalent to the voxelwise variability observed within regions in the measured data \citep{JianAW:09}. Finally, the data was blurred using a standard Gaussian blurring kernel, with FWHM of 6mm, with the voxels in the image being presumed to be 2 mm $\times$ 2 mm (as this is a 2 dimensional simulation) before time independent Gaussian errors with variance proportional to the averaged signal were added. The proportionality of this last measurement error results from the quasi-Poisson nature of errors resulting from Poisson radioactive decay (the reconstruction renders the errors not strictly Poisson, and as such a Gaussian approximation is used in the simulations).

We will compare four deconvolution strategies, the FPCA and spline based (SP) methods from the 1-D simulations and an additional two based on PET parametric models. The curve-by-curve method for the 1-D simulations was also implemented but found to be considerably worse than either FPCA or SP methods (results not shown), so was not considered further. For completeness, the techniques used are now detailed in full. The first is the standard compartmental model based deconvolution based on first order linear ODEs. Given the non-negativity in the parameter values for the model, this can be solved using non-negative least squares \citep{Lawson74}, known as PET spectral analysis in the PET literature \citep{CunnJ:93}. This analysis will be performed using the standard software DEPICT. \cite{JianAW:09} showed borrowing spatial information can reduce the noise and thus improve the $V_T$ estimates by PET spectral analysis. Therefore, an additional comparison will be made with spectral analysis after the data has been pre-processed (pDEPICT). Similarly the approach of \citet{OSullivan:09} will also be applied to the presmoothed data (with results being worse if presmoothing is not preformed).  Finally, the proposed FPCA methodology will be considered. As $V_T$ is the parameter of interest in the PET study, this will be the target of interest in the simulations.

%
%

\subsubsection{Image Simulation 1: Compartmental Regions}

This first image based simulation was designed to assess the performance of the proposed methodology where a true compartmental structure
\[
C_i(t) = \left(\alpha_{i1} e^{\beta_{i1} t} + \alpha_{i2} e^{\beta_{i2} t} \right) \otimes I(t)
\]
was present everywhere, in this case a two compartmental model. This, of course, favors the DEPICT method, where a compartmental structure is assumed, but given that compartmental models are routinely used in PET analysis, and have proved to be useful models in such cases, it is something that is of interest to assess. The values used in each region, along with its size are given in Table \ref{sim1VD}, where the parameters, $\alpha_{ij}$, $\beta_{ij}$ were chosen to coincide with physiologically plausible parameters values from PET studies.

\begin{table}[H]
\caption{Parameters for first 2-D image simulation.} \label{sim1VD}
\centering
\begin{tabular}{|l|rccccc|}
\hline
Region&Size&$\alpha_{i1}$&$\alpha_{i2}$&$\beta_{i1}$&$\beta_{i2}$&$V_T$\\
\hline
1& 9614&-&-&-&-&0\\
2& 5351 & 0.0060& - & 0.0030& - & 2.00 \\
3& 701 &  0.0040  &  0.0023&   0.0008  &  0.0103 & 4.98 \\
4& 14  & 0.0068   & 0.0009&  0.0007  &  0.0203& 9.24 \\
5& 704  & 0.0007 & - & 0.0377 & - & 0.02\\

\hline
\end{tabular}
\end{table}

When comparing the MSE of $V_T$ estimates, Table \ref{MSEISE_sim1} indicates that the FPCA approach and pDEPICT outperform standard spectral analysis (DEPICT) in all five regions even though the data are generated from compartment models. pDEPICT performs better in regions 2, 3 and 4 while FPCA performs better in the rest of the regions. These findings are not too surprising as the data are generated from compartment models which are in favor of the DEPICT approach and \cite{JianAW:09} also has showed borrowing spatial information to reconstruct the signals can further improve the $V_T$ estimates by DEPICT due to noise reduction. However, as can also be seen, a completely non-parametric FPCA approach is still competitive even in this situation where it is possible to assume the correct model structure. However, as expected, the non-parametric approach (SP) which does not involve FPCs performs very badly due to the high noise levels in a voxelwise analysis. In particular, the SP performance is often reasonable, but occasionally has issues at the boundaries (as was seen in Figure \ref{sim0_fig}), which can yield large values of MSE.

\begin{table}[H]
\caption{Averaged MSE (s.e.) of $V_T$'s
based on 50 runs for five different regions in first 2-D simulation} \label{MSEISE_sim1}
\begin{center}\footnotesize
\begin{tabular}{c|ccccc}\hline\hline
 Region &  1 & 2 & 3 & 4 & 5 \\ \hline
  \multirow{2}{*}{FPCA }     &  0.0114  &  0.0832  &  0.4943  &  0.6228  &  0.0433  \\
                           &  (0.0012)  &  (0.0109)  &  (0.1471)  &  (0.8306)  &  (0.0076) \\ \hline
 \multirow{2}{*}{DEPICT}         & 0.1306  &  0.2204  &  0.5549  &  0.6730  &  0.1741  \\
                           &  (0.0113)  &  (0.0076) &   (0.0507)  &  (0.4069)  &  (0.0149) \\ \hline
\multirow{2}{*}{pDEPICT}               &  0.0248  &  0.0601  &  0.2335  &  0.2594  &  0.0505\\
                           & (0.0033)  &  (0.0049)  &  (0.0481)  &  (0.2238)  &  (0.0086) \\ \hline
\multirow{2}{*}{SP}        &  0.0155   &   1.1710   & 7.7870  & 19.5207  &  0.2936 \\
                           &  (0.0014) &   (0.0849) &   (0.4739) &   (4.4023) &   0.0424 \\ \hline\hline
\end{tabular}
\end{center}
\end{table}

\subsubsection{Image Simulation 2: Non-compartmental regions}

The purpose of this second simulation is to investigate how these four approaches perform when the IRFs are not generated from compartment models. Again, we use the brain phantom image with five different regions; however, we replace the IRFs in regions 2 and 4 with scaled mixture gamma pdf's while the other three regions remain the same, thus incorporating a mixture of both compartmental and non-compartmental regions in the simulation. The level of $V_T$ randomness in these two regions is again taken to be are around $6.5\%$. The blurring procedure in the final step is identical to simulation 1. Here is the scaled mixture gamma pdf for region 2,
\begin{equation}\label{eq_MB1}
M_i(t) = \frac{1}{200}\left[1-\int_0^t \{0.7f_{i1}(u/60)+0.3f_{i2}(u/60)\}du \right],
\end{equation}
where $f_{i1}(t)$ and $f_{i2}(t)$ are gamma pdf's with parameters $(\alpha_{i1},\beta_{i1})$ and $(\alpha_{i2},\beta_{i2})$ and $\alpha_{i1}\sim N(1.5,.05^2)$, $\alpha_{i2} \sim N(10,.5^2)$, $\beta_{i1}\sim N(2,.2^2)$ and $\beta_{i2} \sim N(1.5,.1^2)$. The fraction $\frac{1}{200}$ is to make the integral of $M_i(t)$ close to a real value $(\approx 1.97)$. In region 4, the scaled mixture gamma pdf is
\begin{equation}\label{eq_MB2}
M_i(t) = \frac{1}{70}\left[ 1-\int_0^t \{0.8f_{i1}(u/60)+0.2f_{i2}(u/60)\}du \right],
\end{equation}
where $f_{i1}(t)$ and $f_{i2}(t)$ are gamma pdf's with parameters $(\alpha_{i1},\beta_{i1})$ and $(\alpha_{i2},\beta_{i2})$ and $\alpha_{i1}\sim N(2,.15^2)$, $\alpha_{i2} \sim N(15,.1^2)$, $\beta_{i1}\sim N(2.5,.2^2)$ and $\beta_{i2} \sim N(2,.15^2)$.  The fraction $\frac{1}{70}$ is to make the integral of $M_i(t)$ close to a real value $(\approx 8.54)$. The parameters are provided in Table \ref{sim2VD}. The IRF of region 4 has a marked deviation from a compartmental (sum of exponential) structure, while region 2 much more closely resembles a traditional exponential decay, even though it is in fact not expressible as such.

\begin{table}[H]
\caption{Parameters for second 2-D image simulation.} \label{sim2VD}
\centering
\begin{tabular}{|l|rccccc|}
\hline
Region&Size&$\alpha_{i1}$&$\alpha_{i2}$&$\beta_{i1}$&$\beta_{i2}$&$V_T$\\
\hline
1& 9614&-&-&-&-&0\\
2& 5351 & \multicolumn{4}{c}{Mixture Gamma (\ref{eq_MB1})} & 1.97 \\
3& 701 &  0.0040  &  0.0023&   0.0008  &  0.0103 & 4.98 \\
4& 14  & \multicolumn{4}{c}{Mixture Gamma (\ref{eq_MB2})} & 8.54 \\
5& 704  & 0.0007 & - & 0.0377 & - & 0.02\\
\hline
\end{tabular}
\end{table}

Table \ref{MSEISE_sim2} shows the MSE of $V_T$ estimates of the three approaches. As in simulation 1, the FPCA approach outperforms DEPICT in all five regions. pDEPICT outperforms DEPICT except in region 4 and the FPCA approach outperforms pDEPICT in regions 1, 4 and 5. This simulation shows that the preprocessing procedure carried out in pDEPICT can help the DEPICT approach to improve the $V_T$ estimates when compartmental conditions are satisfied; however, it does not always work well. If the true IRFs are close to the assumed compartmental structure, as in region 2, the gains from a model based deconvolution can outweigh the model misspecification errors. However, in situations, such as in region 4, where the true IRF is markedly different from a compartmental structure,  borrowing spatial information from neighboring voxels can result in worse estimates not only against the non-parametric FPCA deconvolution, but even against the standard DEPICT result where no spatial information is taken into account. On the contrary, the FPCA approach estimates the IRFs non-parametrically and thus the performance is more robust and relatively stable regardless of the model structure. Again, a voxelwise non-parametric deconvolution strategy is not competitive, with SP again suffering from large discrepancies in a few of the simulation runs, resulting in very large MSE values overall.

\begin{table}[H]
\caption{Averaged MSE (s.e.) of $V_T$'s
based on 50 runs for five different regions in second 2-D simulation.}\label{MSEISE_sim2}
\begin{center}\footnotesize
\begin{tabular}{c|ccccc}\hline\hline
Region &  1 & 2 & 3 & 4 & 5 \\ \hline
 \multirow{2}{*}{FPCA }          &  0.0116  &  0.0968  &  0.3695  &  0.7045  &  0.0429  \\
        &  (0.0013)  &  (0.0111)  &  (0.1293)  &  (0.4041)  &  (0.0063) \\ \hline
 \multirow{2}{*}{DEPICT}         &  0.1325 &   0.2405  &  0.5545  &  0.9142  &  0.1774  \\
        & (0.0105)  &  (0.0106)  &  (0.0496)  &  (0.4033)  &  (0.0170)  \\ \hline
 \multirow{2}{*}{pDEPICT}    & 0.0257  &  0.0747  &  0.2475  &  1.1158  &  0.0529 \\
                & (0.0028) &   (0.0059)  &  (0.0491)  &  (0.5028)  &  (0.0087) \\ \hline
 \multirow{2}{*}{SP} & 0.0152  &  1.1346   & 7.9536  & 22.8747  &  0.2917 \\
                 & (0.0014) &   (0.0894) &   (0.5365)  &  (5.3599)  &  (0.0471) \\ \hline\hline
\end{tabular}
\end{center}
\end{table}

\begin{figure}[h]
\caption{MSE of the different methods in the regions which are not compartmental models in the 2-D image simulations.}\label{sim2_mse}
\begin{center}
\includegraphics[scale=0.5]{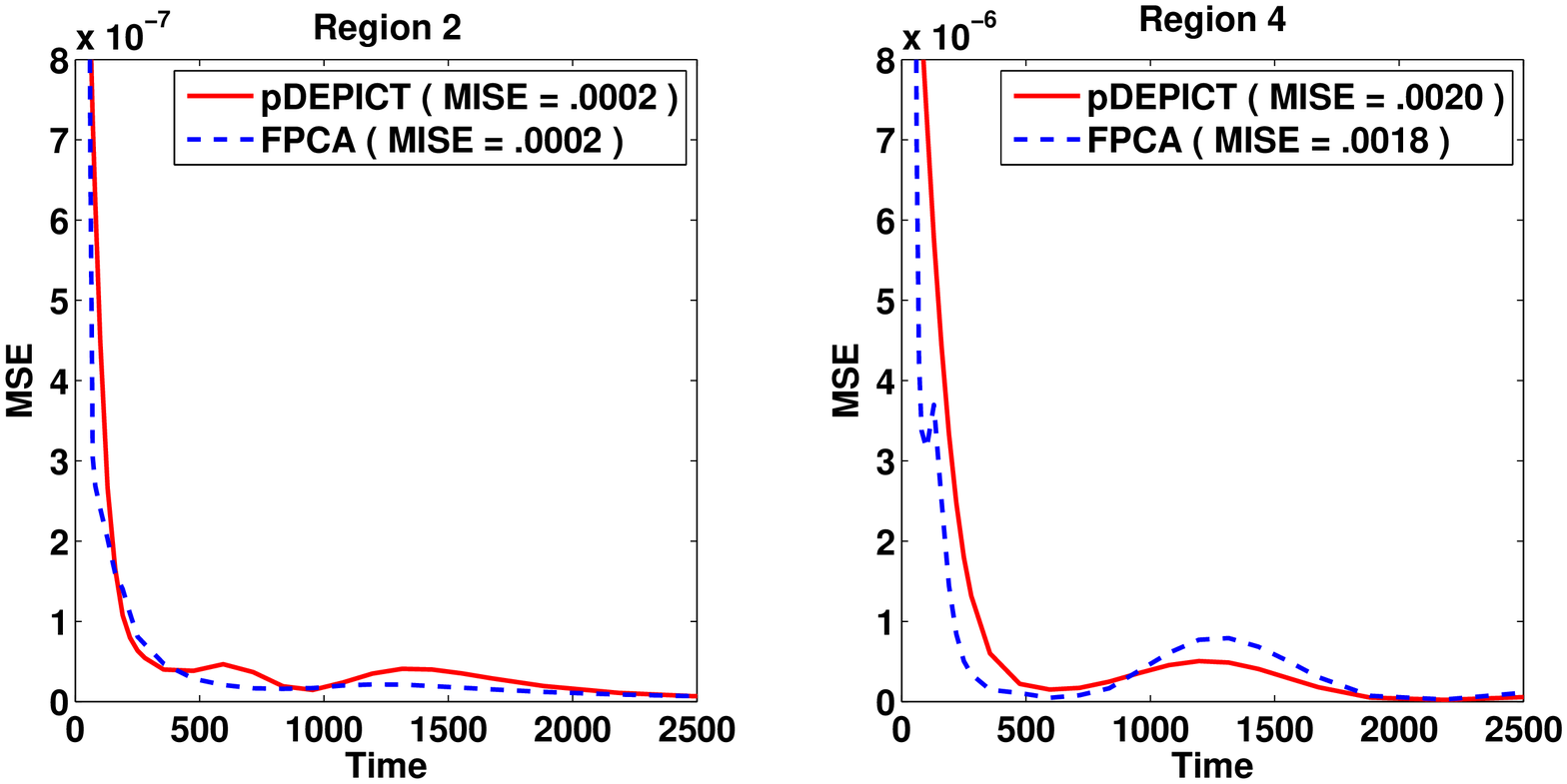}
\end{center}
\end{figure}

From the MSE of the estimated functions in region 2 and particularly in region 4, we see that the FPCA approach captures the function shape nicely while DEPICT can not do so due to its parametric model restrictions. This is emphasised in Figure \ref{sim2_mse} which examines the pointwise MSE of the reconstructed curves in Regions 2 and 4, as well as the MISE. It should be noted at this point though that simply using MISE as a target in this case would indicate that both approaches perform similarly. However, as $V_T$ is the primary interest, we focused on this, and as can be seen in the Table \ref{MSEISE_sim2}, there is a large improvement in MSE for $V_T$ in Region 4 using FPCA.


\section{Measured $^{11}$C-diprenorphine data}\label{s:data}

We apply the nonparametric FPCA approach to a set of dynamic PET scans from a measured [$^{11}$C]-diprenorphine study of normal subjects, for which an arterial input function is available. The main purpose of the study is to produce a population of normal controls to build an understanding of opioid receptor densities in normal brain. Multiple subjects were scanned, some once, some twice. We will analyse this data and focus particularly on the repeated scan data, as this analysis will aim to ensure that the methodology is applicable across a range of subjects with reasonable test-retest reproducibility. While reproducibility cannot be equated with the truth, it is somewhat reassuring if the methods yield relatively similar estimates on the same subject in repeated scans. The scans which are analysed here are part of a study into the relationship between opioid receptors and Epilepsy, and the subjects here are from a normal population for the quantification of opioid receptor distribution. In particular, accurate quantification of $V_T$ is required as this is a measure related directly to receptor density. In addition, it is well known that for the tracer [$^{11}$C]-diprenorphine, any particularly compartmental model does not easily fit the data for all voxels \citep{hammers:2007}, so the investigation of a non-parametric approach is of particular relevance here.

The description of the data here follows from \cite{JianAW:09}, although in that paper only one subject was considered, but the rest were similarly acquired. Each normal control subject underwent either one or two 95-min dynamic [$^{11}$C]-diprenorphine PET baseline scans. The subject was injected with ~185 MBq of [$^{11}$C]-diprenorphine. PET scans were acquired in 3D mode on a Siemens/CTI ECAT EXACT3D PET camera, with a spatial resolution after image reconstruction of approximately 5 mm. Data were reconstructed using the reprojection algorithm \citep{KinaR:89} with ramp and Colsher filters cutoff at Nyquist frequency. Reconstructed voxel sizes were 2.096 mm $\times$ 2.096 mm $\times$ 2.43 mm.  Acquisition was performed in listmode (event-by-event) and scans were rebinned into 32 time frames of increasing duration. Frame-by-frame movement correction was performed on the dynamic [$^{11}$C]-diprenorphine PET images.

\begin{figure}[h]
\caption{Estimated (deconvolved) mean function for subject 2913}\label{dat_Mean}
\begin{center}
\includegraphics[scale=0.26]{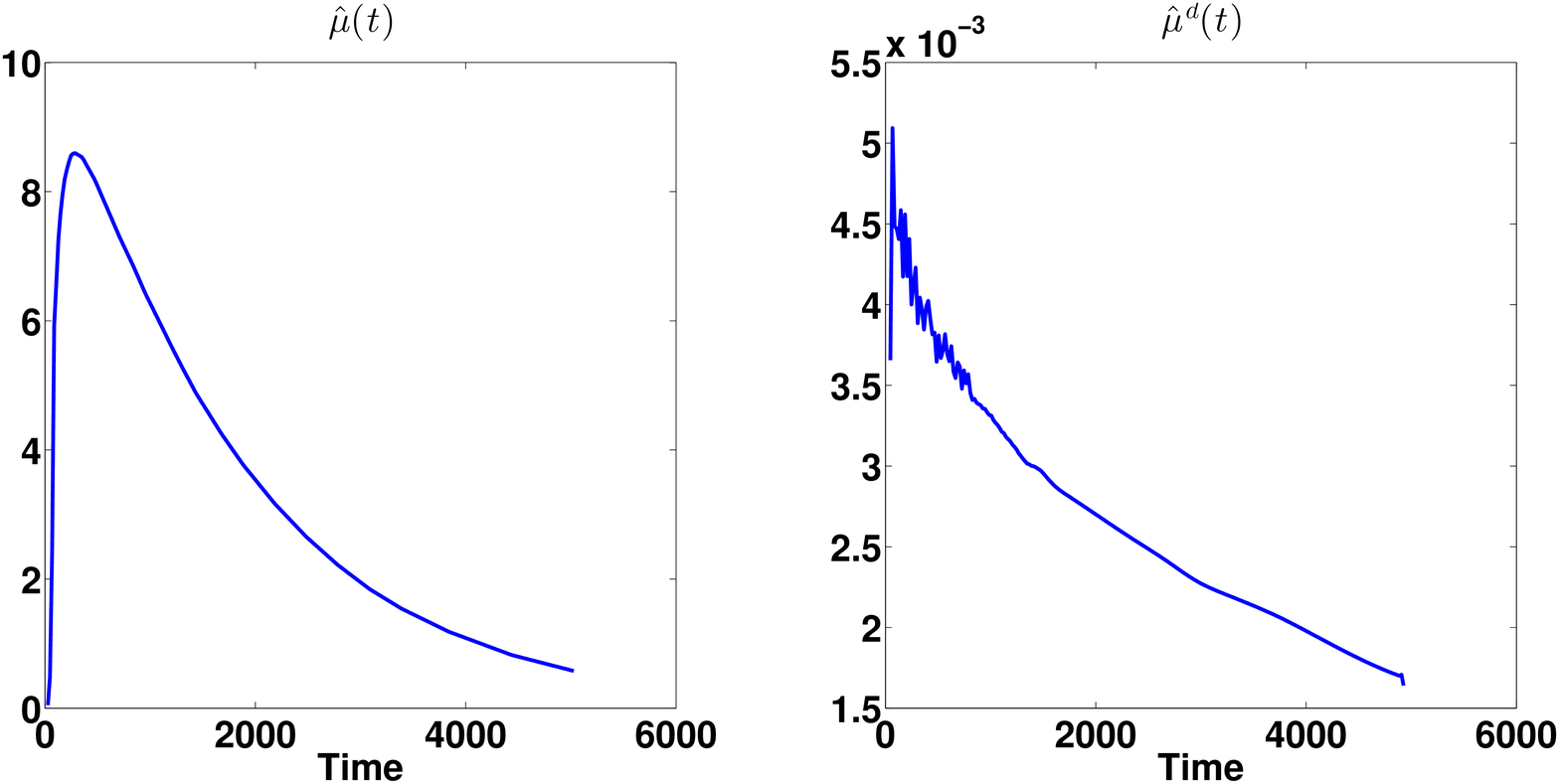}
\end{center}

\caption{The First Three Estimated (deconvolved) Eigenfunctions for subject 2913}\label{dat_Eig}
\begin{center}
\includegraphics[scale=0.26]{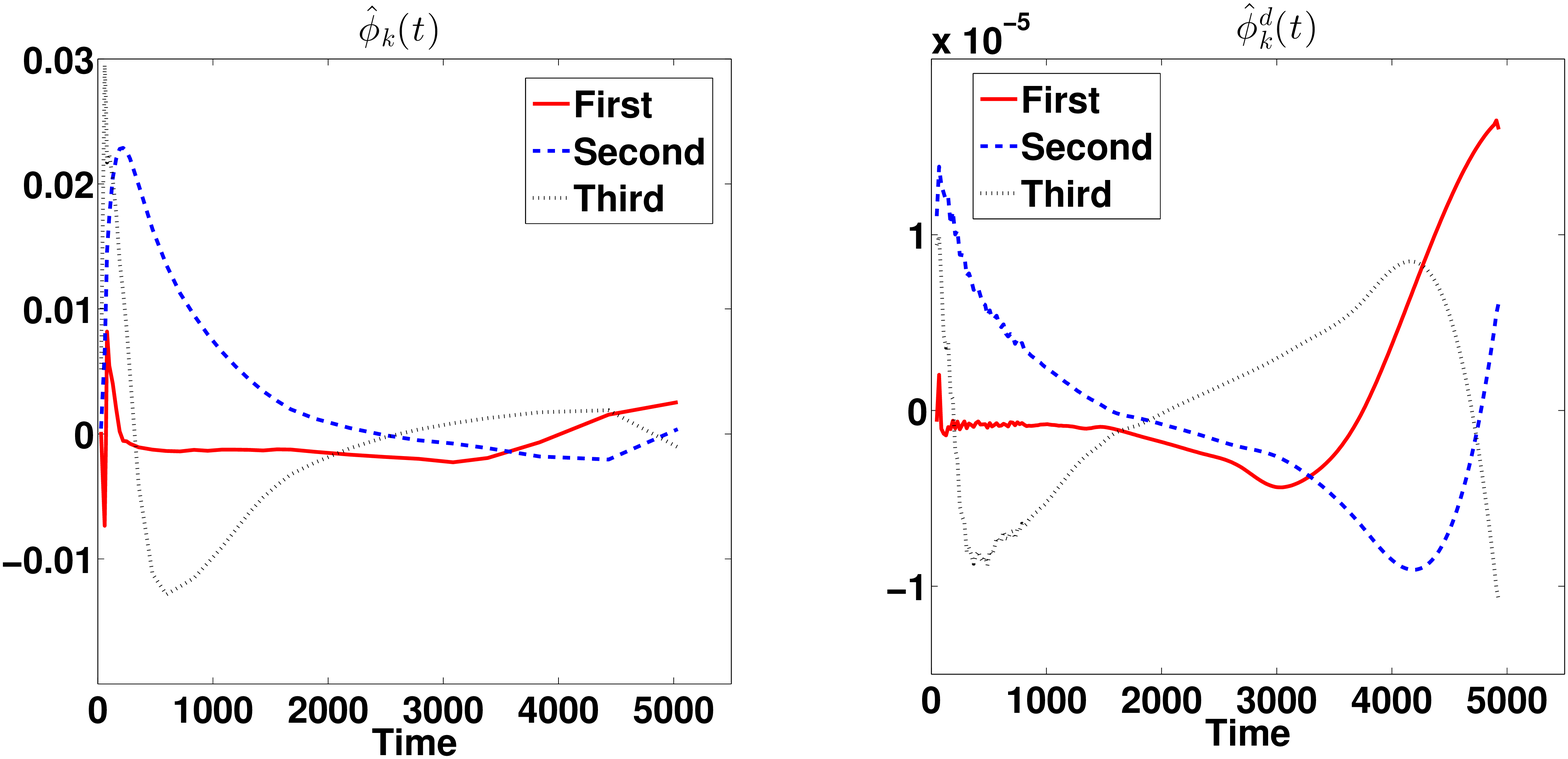}
\end{center}
\end{figure}

The three most promising approaches used in the simulation studies are applied, DEPICT, pDEPICT and the non-parametric FPCA procedure, as the data is on the voxel level, and thus curve by curve methods are both unstable and computationally intractable so not considered further. We first introduce the results of FPCA on a single subject (no. 2913, who only had one scan). Figure \ref{dat_Mean} shows the estimated mean function and its deconvolved function. The deconvolved mean function deviates from the shape which would be expected from a sum of exponential functions. Figure \ref{dat_Eig} shows the first three eigenfunctions together with their corresponding deconvolved functions. The eigenfunctions indicate the variation from the mean function among voxels.

\begin{figure}[h]
\caption{The numbers of components needed to reconstruct the latent signals and the impulse response functions for subject 2913. This indicates that the numbers of components are not randomly distributed in the brain but rather exhibit spatial correlation.}\label{dat_NEig}
\begin{center}
\includegraphics[width=\textwidth]{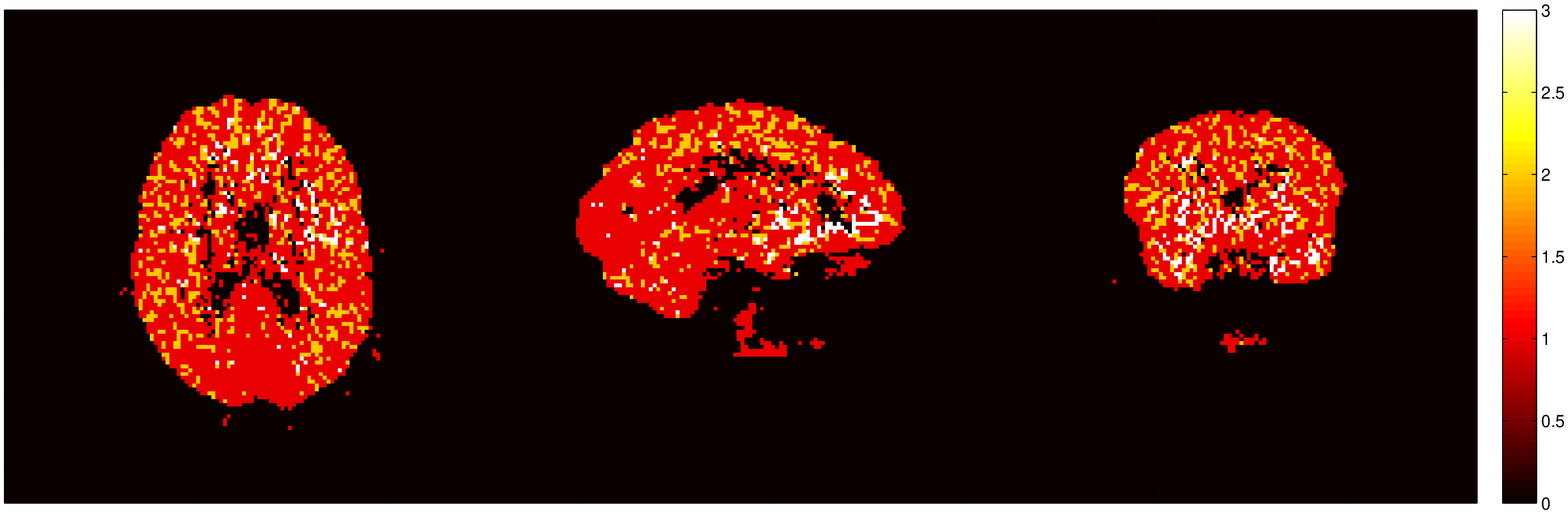}
\end{center}
\end{figure}

\begin{figure}[h]
\caption{The $V_T$ estimates of FPCA approach and DEPICT approach and their differences for subject 2913. The $V_T$s estimated by FPCA are in general smaller than those by DEPICT with $V_T$s reduced about $12.4\%$.}\label{datVD_fpca}
\begin{center}
\includegraphics[width=\textwidth]{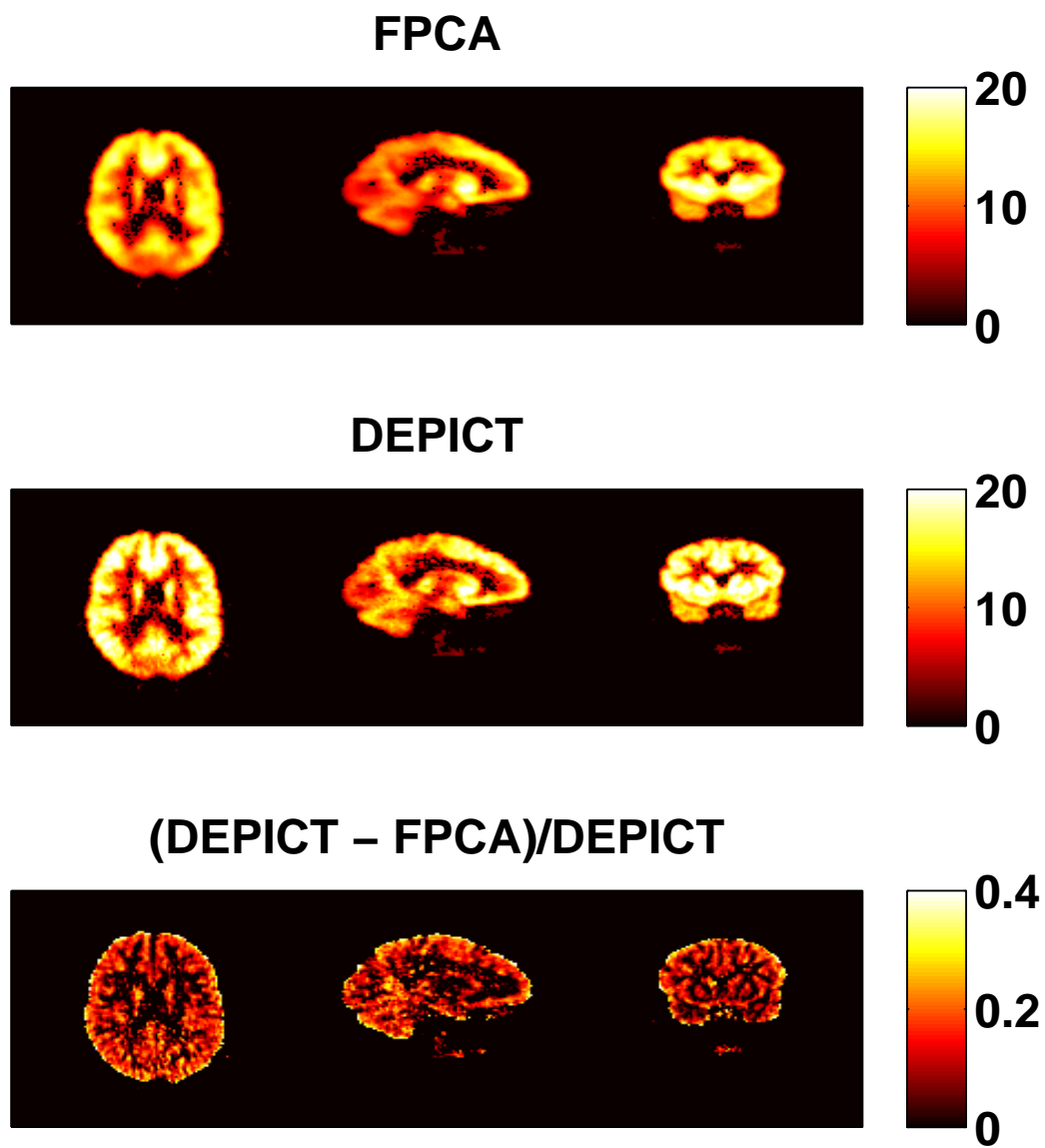}
\end{center}
\end{figure}

Figure \ref{dat_NEig} shows the numbers of components needed to reconstruct the latent signals and the impulse response functions. Spatial clusters exist and it gives indications about the concentration of pain receptors in specific spatial locations. Figure \ref{datVD_fpca} shows the $V_T$ estimates of FPCA approach and DEPICT approach and their differences. The $V_T$ estimates by FPCA are about $12.4\%$ lower than those by DEPICT, while pDEPICT was similar to that of DEPICT (approximately 2.2\% lower, data not shown). Positive biases of 10\% or more are not uncommon for parameter values in the range of those present here (bias is parameter dependent) for PET compartmental models when analyzed with spectral analysis (see \cite{PengAGLA:08}), and as such the estimates provided by FPCA are closer to what might be expected from previous simulation results. Thus, the FPCA yields results which are more quantitatively plausible for comparison across a population. In particular, as differences in PET studies between patients and controls tend to be small, and bias is parameter dependent when using non-linear models, plausible quantitative estimates, which do not rely on particular compartmental assumptions, would allow greater confidence in differences found.

\begin{table}
\caption{Averaged absolute normalised difference for [$^{11}$C]-diprenorphine data. We consider the following measure $\frac{|\widehat{V_T}_1-\widehat{V_T}_2|}{\widehat{V_T}_2}$ if $\widehat{V_T}_2>\delta$ to evaluate the difference between two experiments. Four different $\delta$'s (5,10,15, and 20) are used.  }\label{tr_diff}
\begin{center}\scriptsize
\begin{tabular}{c|cccc|cccc}\hline\hline
           &     \multicolumn{4}{|c}{FPCA} & \multicolumn{4}{|c}{DEPICT} \\ \cline{2-9}
Experiments  &  5  & 10 & 15 & 20 & 5  & 10 & 15 & 20 \\ \hline
244 v.s. 247           &	.186	(.117)	&	.184	(.082)	&	.176	(.070)	&	.148	(.037)	&	.122	(.121)	&	.086	 (.070)	&	.060	(.047)	&	.049	(.036)	\\
1031 v.s. 1033           &	.131	(.157)	&	.088	(.072)	&	.073	(.055)	&	.055	(.040)	&	.158	(.170)	&	.110	 (.093)	&	.087	(.069)	&	.067	(.053)	\\
1248 v.s. 1258           &	.175	(.111)	&	.166	(.056)	&	.180	(.041)	&	.203	(.034)	&	.144	(.120)	&	.124	 (.073)	&	.147	(.063)	&	.163	(.051)	\\
1680 v.s. 1774           &	.243	(.229)	&	.175	(.149)	&	.104	(.079)	&	.063	(.053)	&	.283	(.260)	&	.181	 (.154)	&	.115	(.111)	&	.320	(.354)	\\
1794 v.s. 1798           &	.238	(.221)	&	.184	(.143)	&	.170	(.084)	&	.262	(.033)	&	.295	(.231)	&	.260	 (.175)	&	.293	(.129)	&	.365	(.132)	\\
3427 v.s. 3497           &	.134	(.130)	&	.093	(.064)	&	.060	(.041)	&	NaN	            &	.166	(.161)	&	.106	 (.084)	&	.094	(.068)	&	.168	(.101)	\\
3568 v.s. 3715           &	.193	(.183)	&	.137	(.110)	&	.094	(.077)	&	.076	(.062)	&	.202	(.191)	&	.132	 (.107)	&	.095	(.076)	&	.099	(.085)	\\ \hline
Pooled & .185	(.175) & 	.146	(.110)&	.113	(.078)&	.080	(.067)&	.196	(.196)	& .142	(.127)&	.121	(.109)	& .081	 (.083) \\ \hline\hline
\end{tabular}
\end{center}
\end{table}

%

\begin{figure}[h]
\caption{The $V_T$ estimates of a test and a retest scan from a single subject who had two scans (numbered 1031 and 1033). In addition the percentage change between the two is given for both DEPICT and the FPCA procedure. The difference is truncated at 40\% as all voxels above this had estimated $V_T$ close to 0.}\label{1031vd_per}
\begin{center}
\includegraphics[width=0.49\textwidth]{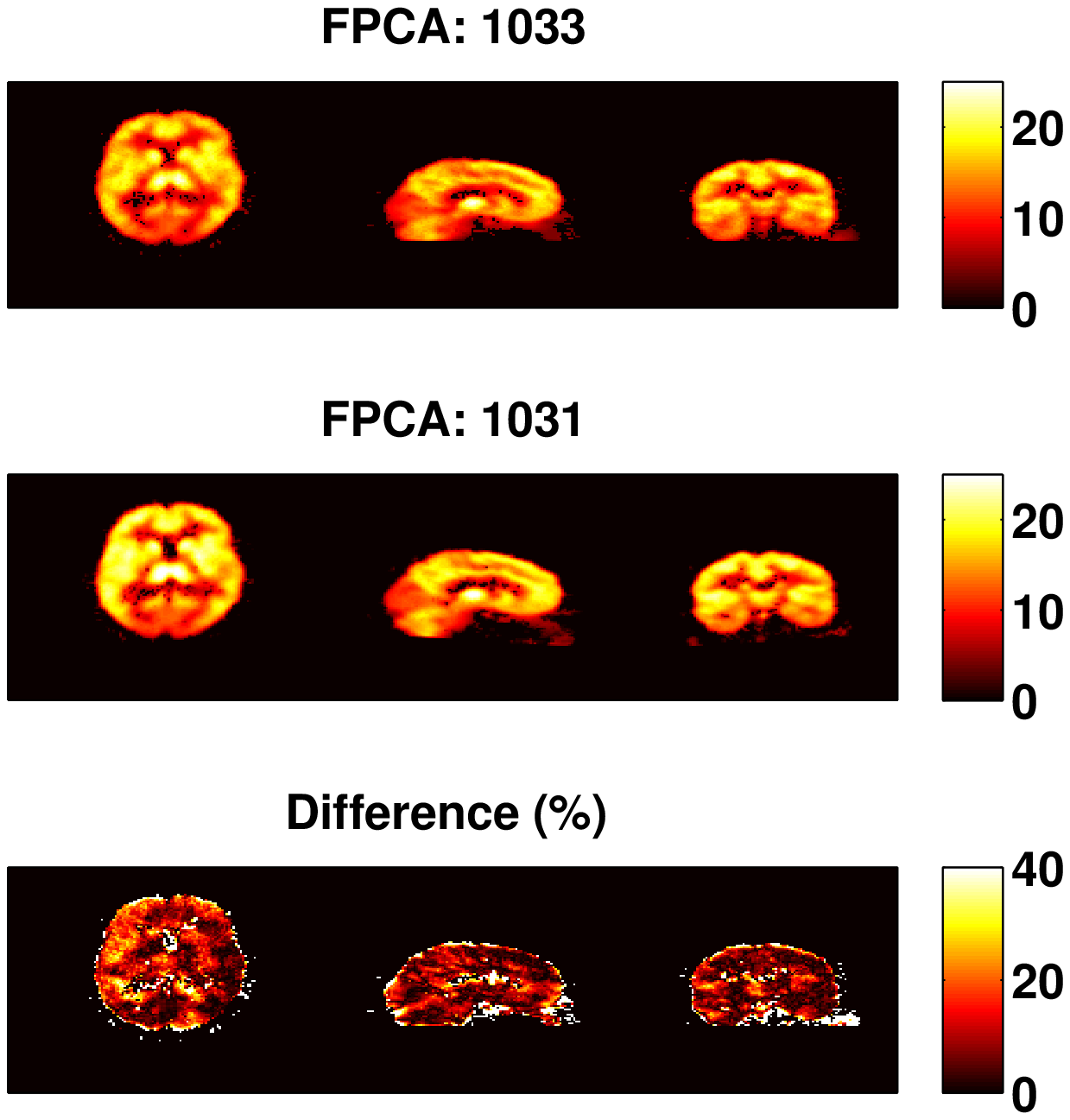}
\includegraphics[width=0.49\textwidth]{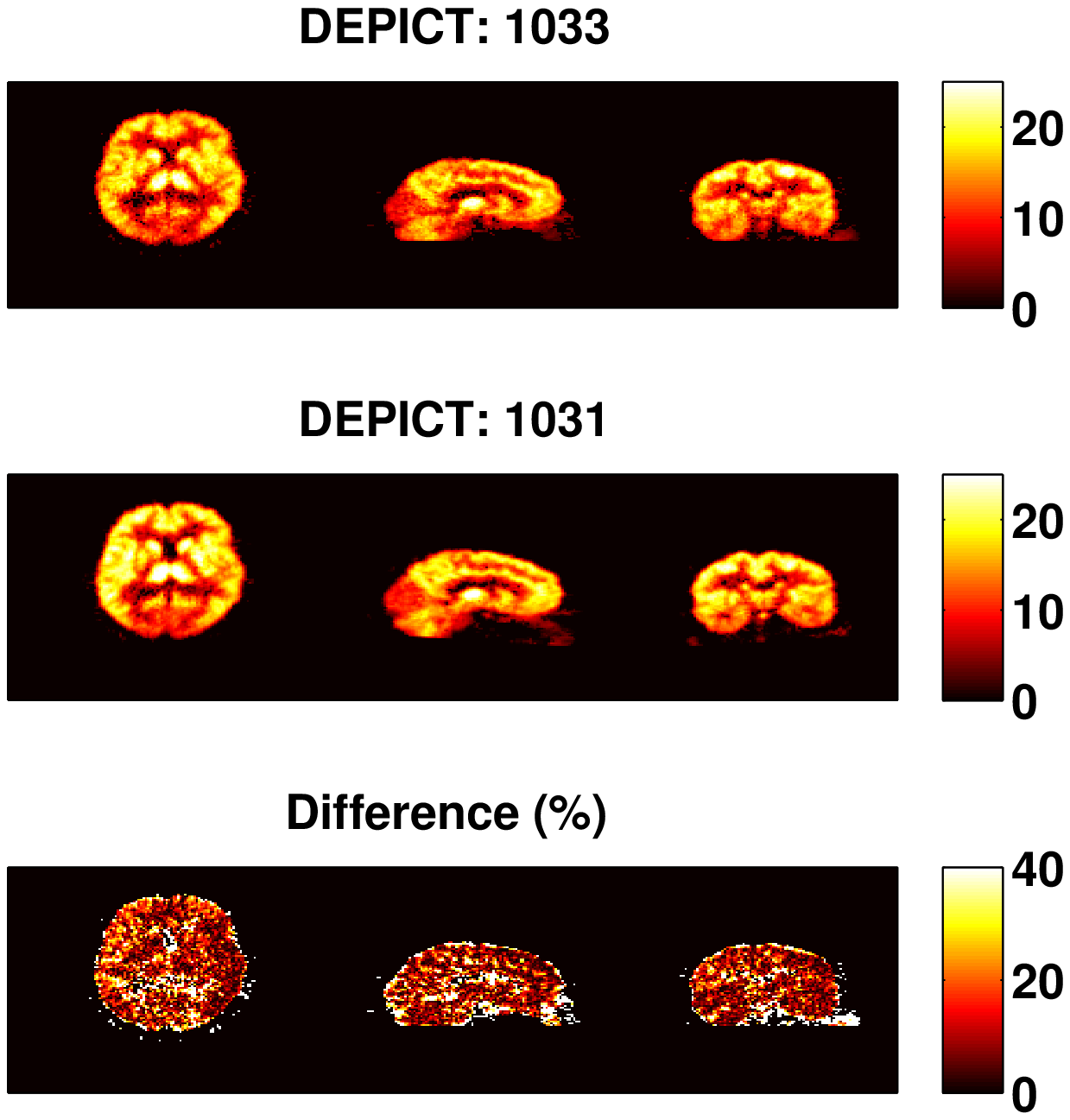}
\end{center}
\end{figure}

The results from the test-retest analysis are given in Table \ref{tr_diff} and Figure \ref{1031vd_per}. Taking the figure first, we see that the test-retest variability of both DEPICT and FPCA are roughly similar in corresponding brain regions. This is reassuring as the FPCA procedure is considerably more flexible than the model based DEPICT estimates. In addition, there is evidence of spatial smoothness in the reproducibility which is physiologically more interpretable from the FPCA approach than the DEPICT approach. Turning to Table \ref{tr_diff}, we see that there is considerable correspondence between the test-retest results from DEPICT and FPCA. The pooled results show that there are similar levels of variation at different threshold levels. This is important, as the receptor densities are only of interest at somewhat higher $V_T$ values. Indeed there is also less variation within the test-retest values for the FPCA procedure. For individual subjects, the results are fairly balanced with some subjects having smaller test-retest differences with FPCA and others with DEPICT at different threshold levels. However, in almost all cases, the variability in the test-retest results is smaller with FPCA than with DEPICT.

From a computational point of view, the time for the non-parametric deconvolution is very competitive to the parametric modeling approach. The proposed procedure took approximately 8.5 mins to analyze a single PET scan, which compares with approximately 10 mins for DEPICT to perform an equivalent analysis (all computations carried out on an Intel core i7 CPU M620 2.67GHz with 4GB RAM).

\section{Discussion}\label{s:disc}

We have presented a functional data analysis approach to the problem of mass deconvolution in neuroimaging. By expressing the deconvolution problem via a functional principal component basis expansion, it is possible to dramatically reduce the required computational complexity. The methodology has been shown to work well both in generic 1-D function deconvolution and also in more realistic image based simulations, while also producing physiologically plausible results in a real data analysis, without resorting to modelling assumptions that are challengeable at best.

The approach to the methodology here has been to take as simple approach as possible for each inherent step. This, of course, could be relaxed, and much more complex algorithms for deconvolution could be investigated in the place of the simple linear deconvolution suggested here. In addition, different methods for choosing the number of eigenfunctions to examine or how the smoothing is performed could also be changed, but without any significant effect on the application of the methodology.

It would be possible to carry out such an analysis using different basis functions using methods such as those explored in \cite{OSullivan:14} for FDG. There a segmentation algorithm is used to determine the basis functions and is shown to work well for FDG. However, when using segmentation algorithms, it is often hard to know how many basis functions to use, particularly for tracers such as [$^{11}$C]PK-11195, a marker for neurodegeneration, which has little spatial coherence, and it is not clear that the resulting decomposition would always be identifiable. However, the eigenbasis approach as proposed here would be equally valid in such a situations and by definition always yields an identifiable basis.

We have here suggested the use of the multiplicative random effects model for the FPCA analysis. This could be replaced with the more usual standard FPCA decomposition. However, it has been shown previously \citep{JianAW:09} that this model is a natural model for PET, given the compartmental assumptions usually made in data modelling, both in terms of its interpretation as well as its empirical performance, and for this reason we have concentrated on it here. It should, of course, be noted that the use of smoothed estimates for the curves, but unsmoothed data to estimate the mean yields the possibility that the functions of the data used to generate the principal component scores will not be eigenfunctions. However, asymptotically (as the smoothing bandwidth goes to zero), these will be consistent estimates. For finite samples, these will still be a completely valid function basis to express the data, albeit not necessarily the finite sample eigenfunctions. However, the gains in using smoothed data to control the noise is considerable over the use of raw curves for deconvolution.

The methodology presented here is naturally appealing for PET data, given that it reduces the number of deconvolutions from several hundred thousand to four or five. However, it is also a candidate for deconvolution for neuroimaging in general, where in modalities such as fMRI, there is interest in deconvolving hemodynamic response functions from the data \citep{Wangetal:13,Zhangetal:12}. A similar FPCA setup to deconvolve fMRI data could therefore be used, although care would need to be taken and additional regularisation used in the deconvolution step, as the null space of the linear operator will be non-zero for fMRI data (due to the negative dip in the hemodynamic response), unlike the case for PET data. Indeed, under suitable assumptions, the approach that has been proposed is applicable in many situations where there are replicates of the curves present, allowing the deconvolution to be treated from a functional data perspective.

%
%
%
%
%
%

\section*{Acknowledgements}
The authors would like to thank Alexander Hammers and Federico Turkheimer for supplying the measured data. 
The research of Ci-Ren Jiang is supported in part by NSC 101-2118-M-001-013-MY2 (Taiwan); the research of Jane-Ling Wang is supported by NSF grants, DMS-09-06813 and DMS-12-28369. JA is supported by EPSRC grant EP/K021672/1. The authors would like to thank SAMSI and the NDA programme where some of this research was carried out.

\appendix
\section{Assumptions}
For notation simplicity, we let $X(t,i)=C_i(t)\exp(-\lambda t)$, $\delta^2(t,i)$ be the variance of $X$ at time $t$ and location $i$, $n$ be the number of voxels, and $N$ be the number of observations per voxel. Suppose the orders of bandwidths are all of the same order as $h$. The following are specific assumptions for the presmoothing step.
\begin{enumerate}
\item The second derivatives of $X(t,i)$, the variable bandwidth function $h_T(t)$ and $\delta^2(t,i)$ are continuous and bounded.

\item The kernel function $K$ assigns weights to each data point and is assumed to a symmetric probability density function with bounded support.

\item For a $k$-dimensional smoother, $h\rightarrow 0$ and $nNh^{k} \rightarrow \infty$. 

\end{enumerate}

In addition, we assume that the input function $I(t)$ is smooth and positive over the entire range of the integration. This is true in practice given the nature of the input function being the amount of tracer in the blood plasma.


\bibliography{fda}
\bibliographystyle{chicago}

\end{document}